\documentclass[journal]{IEEEtran}
\usepackage{cite}
\usepackage{amsmath}
\usepackage{times}
\usepackage{epsfig}
\usepackage{graphicx}
\usepackage{subfig}
\usepackage{amsmath}
\usepackage{amssymb}
\usepackage{tabularray}
\usepackage{tikz}
\usepackage{booktabs}
\usepackage[ruled,vlined]{algorithm2e}  
\usepackage{stmaryrd}
\usepackage{caption}  
\usepackage{longtable} 
\usepackage{float} 
\usepackage{url}
\usepackage{subcaption}
\usepackage{xurl}
\usepackage{afterpage}
\usepackage{comment}
\def\BibTeX{{\rm B\kern-.05em{\sc i\kern-.025em b}\kern-.08em
    T\kern-.1667em\lower.7ex\hbox{E}\kern-.125emX}}
\begin{document}

\title{Predictive Beamforming in Low-Altitude Wireless Networks: A Cross-Attention Approach}

\author{Xiaotong Zhao, Yuanhao Cui, Weijie Yuan, Ziye Jia, Heng Liu, Chengwen Xing
    \thanks{
    Xiaotong Zhao and Yuanhao Cui are with the School of Information and Communication Engineering, Beijing University of Posts and Telecommunications, Beijing 100876, China (xiaotongzhao@bupt.edu.cn; yuanhao.cui@bupt.edu.cn).
    
    Weijie Yuan is with the School of System Design and Intelligent Manufacturing, Southern University of Science and Technology, Shenzhen 518055, China (yuanwj@sustech.edu.cn).
    
    Ziye Jia is with the College of Electronic and Information Engineering, Nanjing University of Aeronautics and Astronautics, Nanjing 211106, China, and also with the National Mobile Communications Research Laboratory, Southeast University, Nanjing, Jiangsu, 211111, China (e-mail: jiaziye@nuaa.edu.cn).
    
    Heng Liu and Chengwen Xing are with the School of Information and Electronics, Beijing Institute of Technology, Beijing 100081, China (heng\_liu\_bit\_ee@163.com; chengwenxing@bit.edu.cn).

    The corresponding author is Weijie Yuan.
    }
}
\maketitle
\pagestyle{empty}
\thispagestyle{empty}
\begin{abstract}
Accurate beam prediction is essential for maintaining reliable links and high spectral efficiency in dynamic low-altitude wireless networks. However, existing approaches often fail to capture the deep correlations across heterogeneous sensing modalities, limiting their adaptability in complex three-dimensional environments. To overcome these challenges, we propose a multi-modal predictive beamforming method based on a cross-attention fusion mechanism that jointly leverages visual and structured sensor data. The proposed model utilizes a Convolutional Neural Network (CNN) to learn multi-scale spatial feature hierarchies from visual images and a Transformer encoder to capture cross-dimensional dependencies within sensor data. Then, a cross-attention fusion module is introduced to integrate complementary information between the two modalities, generating a unified and discriminative representation for accurate beam prediction. Through experimental evaluations conducted on a real-world dataset, our method reaches 79.7\% Top-1 accuracy and 99.3\% Top-3 accuracy, surpassing the 3D ResNet–Transformer baseline by 4.4\%–23.2\% across Top-1 to Top-5 metrics. These results verify that multi-modal cross-attention fusion is effective for intelligent beam selection in dynamic low-altitude wireless networks.
\end{abstract}
\section{Introduction}
With the evolution of 6G communication towards space-air-ground integrated networks, low-altitude airspace has become a key operational domain that supports emerging applications such as urban logistics, emergency response, and public safety \cite{low1, low2, low3, low4, low5}. However, even in sub-6 GHz frequency bands, low-altitude Unmanned Aerial Vehicle (UAV) communication systems suffer from severe beam mismatch due to the highly dynamic and three-dimensional nature of air-to-ground links. Rapid variations in the UAV’s position, altitude, and orientation continuously alter the propagation geometry, resulting in beam misalignment and unstable link gains. Moreover, in dense urban environments, complex multipath propagation and dynamic obstacles further aggravate the problem, causing irregular signal reflections, intermittent blockage, and severe degradation in effective communication capacity.

To address path loss and beam misalignment, conventional cellular systems employ exhaustive beam training where the base station scans through a predefined codebook and selects the beam featuring the strongest received power. While effective in static scenarios, this approach suffers from substantial training overhead that consumes communication resources and reduces data transmission time. It also lacks environmental awareness, potentially selecting blocked beams, and exhibits poor adaptability to rapid UAV positional and attitudinal changes in dynamic 3D flight environments, leading to frequent beam mismatch and link quality degradation.

Recently, research efforts have shifted toward sensing-aided beam prediction to overcome these limitations \cite{beam2, zhao2025fl, beam3}. This class of methods leverages sensing data to learn a mapping between environmental or positional information and the optimal beam index, thereby reducing training overhead and enhancing beam alignment accuracy. For instance, Morais \textit{et al.} \cite{morais2023position} and Heng \textit{et al.} \cite{heng2021machine} exploited positional data to assist beam selection, correlating UAV location with beam direction to narrow the search space. Similarly, Wang \textit{et al.} \cite{wang2024vision} and Xiang \textit{et al.} \cite{xiang2023computer} employed visual data to infer the existence of obstacles and environmental structures, thereby avoiding blocked or ineffective beam paths. Although these single-modality methods have successfully reduced training cost and improved robustness against occlusions, their limited sensing dimension constrains their adaptability in highly dynamic and cluttered low-altitude environments.

This paper adopts a multi-modal sensing-based predictive beamforming method for low-altitude UAV communication that fuses visual, positional, and inertial data to infer the optimal beam without traditional beam search. Specifically, we design a beam prediction architecture based on cross-attention, in which visual features are extracted from a CNN, structured sensor features by a Transformer, and then fused by cross-attention to adaptively align multimodal information. The proposed method is validated based on real-world datasets, e.g., DeepSense 6G, showing that it achieves strong Top-1 and Top-3 accuracies in realistic, dynamic low-altitude settings.

The key contributions of this paper are outlined as follows:
\begin{enumerate}
    \item We propose a multi-modal sensing-based predictive beamforming method for low-altitude wireless communication. By leveraging environmental and UAV motion information, the method enables rapid beam alignment without the need for exhaustive beam search, significantly reducing training overhead and latency.
    
    \item We design a cross-attention-based beam prediction framework that combines CNN and Transformer modules for data preprocessing and feature extraction, and employs a cross-attention mechanism for multimodal fusion. This design allows effective interaction between visual and structured sensing modalities, improving prediction accuracy and robustness.
    
    \item We validate the proposed method using a real-world dataset from DeepSense 6G. The results show that the model achieves superior Top-1 and Top-3 accuracies compared with existing methods, demonstrating its effectiveness for fast and reliable beam prediction in dynamic low-altitude environments.
\end{enumerate}

The remaining sections of this paper are arranged as follows. Section I introduces related works. Section II develops the system model and problem formulation. Section III details the proposed CNN–Transformer cross-attention beam prediction architecture. Section IV evaluates performance on real-world data. Section V concludes and discusses future work.

\section{System Model and Problem Formulation}
\subsection{System Model}
\begin{figure}[t]
    \centering
    \includegraphics[width=0.8\linewidth]{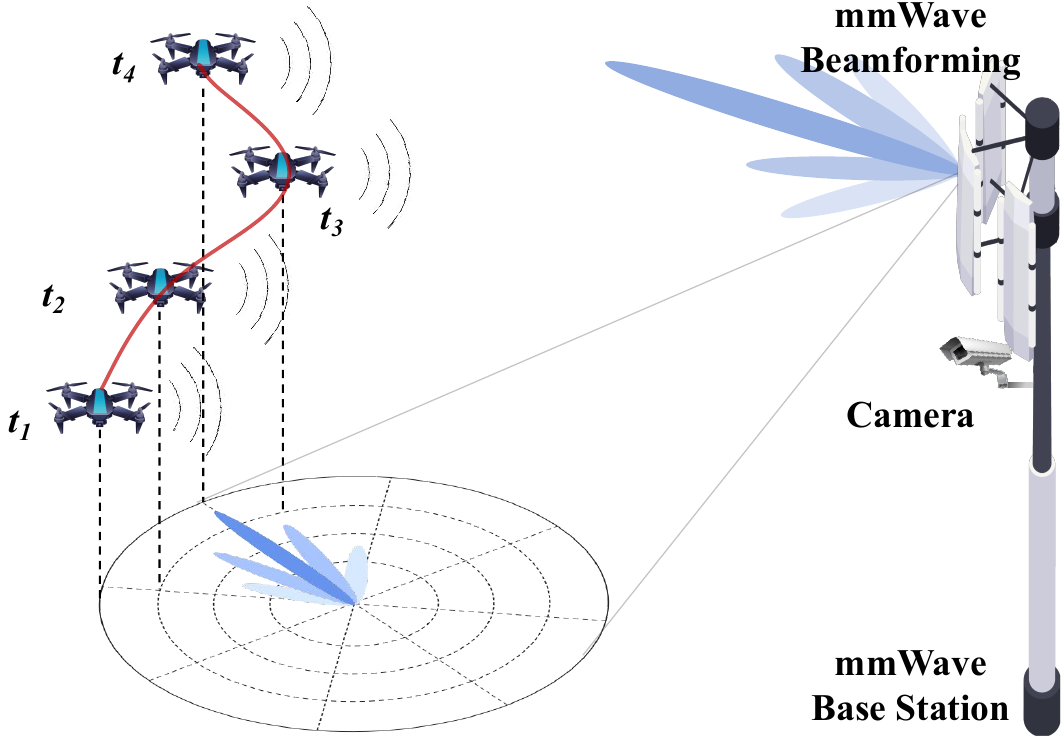}
    \caption{Multi-modal sensing-aided communication scenario.}
    \label{fig:1}
\end{figure}

The system model, illustrated in Fig.~\ref{fig:1}, comprises a Base Station (BS) with an $M$-element uniform linear array (ULA) and an RGB camera, while a UAV is equipped with a single-antenna transmitter and onboard sensor. The UAV’s GPS receiver and Inertial Measurement Units (IMU) jointly provide real-time position and motion information (latitude, longitude, altitude, velocity, and attitude), which are later utilized as auxiliary sensing inputs.

For the downlink, communication is implemented using orthogonal frequency-division multiplexing (OFDM), configured with $K$ subcarriers, effectively mitigating multipath delay and Doppler effects. To support directional transmission, the BS maintains $\mathcal{F} = \{\mathbf{f}_q\}_{q=1}^{Q}$ (a beamforming codebook ), where each beam vector $\mathbf{f}_q \in \mathbb{C}^{M \times 1}$ denotes a candidate transmission direction.

The channel vector of the $k$-th subcarrier at time $t$ is denoted as \(\mathbf{h}_k[t] \in \mathbb{C}^{M \times 1}\). The baseband signal that the UAV receives is:
\begin{equation}
    y_k[t] = \mathbf{h}_k^T[t]\mathbf{f}_q[t]x + v_k[t],
\end{equation}
in the above expression, \(\mathbf{f}_q[t] \in \mathcal{F}\) represents the selected beamforming vector, $x$ denotes the transmitted symbol characterized by the condition \(\mathbb{E}[|x|^2]=P\), and \(v_k[t]\sim\mathcal{N}_\mathbb{C}(0,\sigma^2)\) signifies complex Gaussian noise.

To enhance the communication quality, the most appropriate beamforming vector \(\mathbf{f}^\ast[t]\) is picked from the codebook via maximizing the mean subcarrier gain:
\begin{equation}
    \mathbf{f}^\ast[t] =
    \arg\max_{\mathbf{f}_q[t]\in\mathcal{F}}
    \frac{1}{K}\sum_{k=1}^{K}
    \left|\mathbf{h}_k^T[t]\mathbf{f}_q[t]\right|^2
    \cdot \frac{P}{\sigma^2}.
    \label{eq:2}
\end{equation}
where \(P\) denotes the transmit power and \(\sigma^2\) is the noise variance. The term thus scales the raw channel gain to reflect the effective signal-to-noise ratio (SNR) at the receiver. This formulation describes the spatial filtering process performed by the BS to align the transmission beam with the UAV’s dominant propagation direction, forming the foundation for subsequent sensing-aided beam prediction.
\subsection{Problem Formulation}
\begin{figure*}[t]
    \centering
    \includegraphics[width=0.85\textwidth]{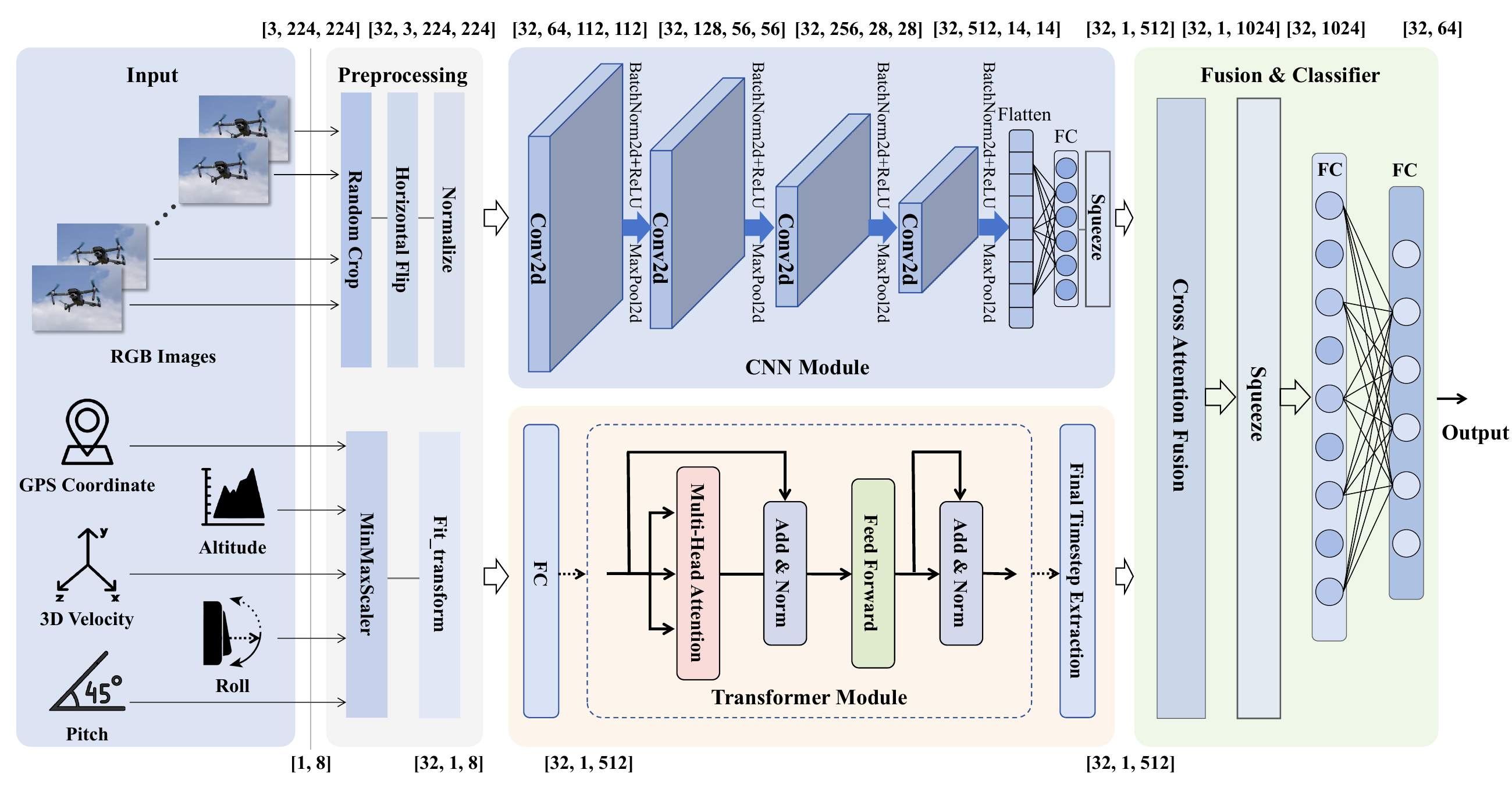}
    \caption{Architecture of the proposed cross-attention CNN–Transformer fusion network for multimodal beam prediction.}
    \label{fig:2}
\end{figure*}
Based on the system model outlined in Section~II, the objective of beam selection for mmWave/THz UAV communication lies in determining the optimal beamforming vector \(\mathbf{f}^\ast[t]\) which maximizes the received SNR specified in (\ref{eq:2}). However, explicit channel estimation and exhaustive beam search impose substantial training overhead and are infeasible in dynamic UAV environments. To tackle this constraint, we redefine the beam selection problem into a multimodal beam prediction task, in which sensing data serve to deduce \(\mathbf{f}^\ast[t]\) directly without the requirement of explicit CSI acquisition.

At each time instant $t$, the multimodal sensing input is specified as:
\begin{equation}
    \mathcal{S}[t] = \{\mathbf{X}[t], \mathbf{P}[t], \mathbf{M}[t]\},
\end{equation}
where $\mathbf{X}[t] \in \mathbb{R}^{W \times H \times 3}$ denotes the RGB image acquired via the BS RGB camera, $\mathbf{P}[t] \in \mathbb{R}^3$ represents position data (latitude, longitude, and altitude), and $\mathbf{M}[t] \in \mathbb{R}^5$ contains motion–attitude information (3D velocity and two attitude angles). The structured data $\{\mathbf{P}[t], \mathbf{M}[t]\}$ thus form an 8-dimensional auxiliary feature vector complementary to visual information.

The learning objective is to establish a mapping between the multimodal sensing data and the optimal beamforming vector:
\begin{equation}
    f_{\Theta}: \mathcal{S}[t] \rightarrow \mathbf{f}^\ast[t],
\end{equation}
where $\Theta$ represents the learnable model parameters. In practice, the model determines the index associated with the beamforming vector \(\mathbf{f}^\ast[t]\) inside the predefined codebook \(\mathcal{F} = \{\mathbf{f}_q\}_{q=1}^{Q}\), i.e.,
\begin{equation}
    f_{\Theta}: \mathcal{S}[t] \rightarrow \hat{q}[t],
    \quad \mathbf{f}^\ast[t] = \mathbf{f}_{\hat{q}[t]}.
\end{equation}
The objective is to minimize the discrepancy between $\mathbf{f}^\ast[t]$ and the true optimal beam obtained via (\ref{eq:2}), thereby achieving accurate and low-overhead beam alignment in dynamic UAV communication scenarios.

\section{Proposed Multi-modal Beam Prediction: A CNN–Transformer Cross-Attention Approach}

As shown in Fig.~\ref{fig:2}, the proposed model is designed to predict the optimal beamforming vector $\mathbf{f}^\ast[t]$ based on multimodal sensing data. The overall framework consists of four core modules: (1) multimodal data preprocessing, (2) CNN-based visual feature extraction, (3) Transformer-based structured feature extraction, and (4) cross-attention fusion and classification. The design and implementation of each component are described below.
\subsection{Multimodal Data Preprocessing}
\subsubsection{Visual Data Preprocessing}
In low-altitude wireless networks, the visual modality of the base station camera is typically oriented toward the sky to capture the UAV within its field of view. Consequently, the acquired RGB images are often affected by illumination variation, weather dynamics, e.g., cloud cover, fog, or sunlight reflection, and large viewpoint changes caused by UAV mobility. These factors introduce substantial visual diversity and noise, which can degrade the robustness of feature extraction. 

To mitigate these effects and enhance generalization, a series of data augmentation and normalization operations is applied. The preprocessing pipeline adopts a fixed spatial size of \(224 \times 224\) pixels for all input images to maintain dimensional consistency. Then, random horizontal flipping and random cropping with 16-pixel padding are performed to simulate variable camera angles and partial occlusion. Finally, the pixel intensities are scaled to \([0, 1]\) and then standardized by the ImageNet mean and standard deviation:
\begin{equation}
    \hat{\mathbf{X}}[t] = \frac{\mathbf{X}[t] - \mu}{\sigma},
\end{equation}
where $\mu = [0.485, 0.456, 0.406]$ and $\sigma = [0.229, 0.224, 0.225]$. This preprocessing procedure improves illumination invariance, stabilizes the gradient distribution during training, and accelerates model convergence.
\subsubsection{Structured Data Processing}
In addition to visual information, the collected data of UAV position and motion–attitude measurements originate from heterogeneous onboard sensors, each operating under different physical principles and output units. Let $\mathbf{S}[t] = \{\mathbf{P}[t], \mathbf{M}[t]\}$ denote the 8-dimensional structured feature vector comprising position and motion–attitude information. The resulting multimodal data exhibit significant disparities in scale, range, and noise distribution, which can degrade the stability of model training and hinder cross-modal feature fusion.

To ensure the compatibility of heterogeneous sensor data, the structured feature vector $\mathbf{S}[t]$ is normalized using min–max scaling:
\begin{equation}
\hat{\mathbf{S}}[t] = \frac{\mathbf{S}[t] - \min(\mathbf{S})}{\max(\mathbf{S}) - \min(\mathbf{S})}.
\end{equation}
Here, $\min(\mathbf{S})$ and $\max(\mathbf{S})$ denote the minimum and maximum feature values observed in the training set. This normalization ensures that all structured features are projected onto a consistent range of $[0,1]$, which prevents features with large numeric magnitudes (e.g., velocity or altitude) from dominating the learning process. Moreover, by aligning feature scales before entering the classifier, this step facilitates stable convergence and more effective interaction between the structured and visual feature spaces during the subsequent cross-attention fusion stage.
\begin{figure*}[t]
    \centering
    \includegraphics[width=0.85\textwidth]{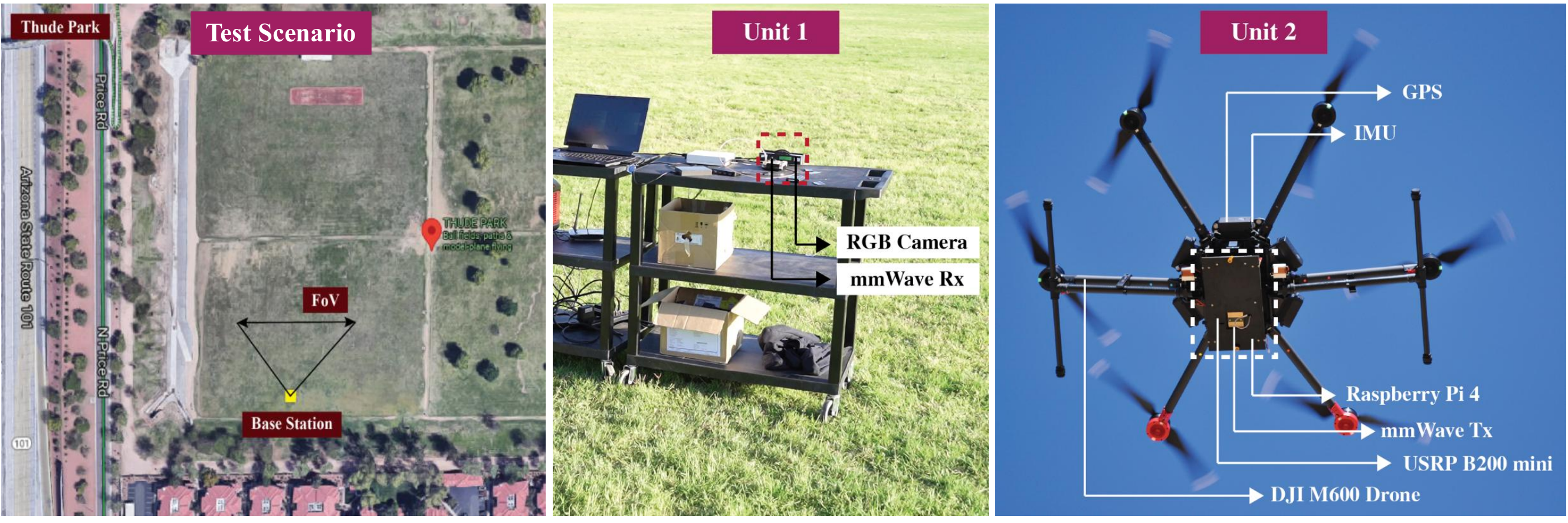}
    \caption{Real-world multimodal data acquisition and testing setup, as shown in \cite{deepsense6G}. The system includes a BS (Unit 1) equipped with an RGB camera and mmWave receiver, and a UAV (Unit 2) carrying GPS, IMU, and mmWave transmitter.}
    \label{fig:3}
\end{figure*}
\subsection{CNN Feature Extraction}
In low-altitude wireless communication, visual data offer rich spatial cues and hierarchical structures from local textures to global semantics. CNNs effectively extract such multi-scale features through local connectivity and weight sharing, which enhance generalization and mitigate overfitting. Their translation invariance further ensures robustness to UAV positional variations within the visual frame.

The visual feature extraction pipeline in this work is built around a four-layer convolutional core. Each convolutional layer is successively followed by batch normalization, a ReLU activation function, and a max-pooling operation. The preprocessed RGB image input is represented as a tensor of size $[3, 224, 224]$. The initial convolutional layer uses 64 filters sized $3 \times 3$, resulting in a feature map of $[64, 224, 224]$, followed by $2 \times 2$ max pooling that downsamples it to $[64, 112, 112]$ to reduce computation while retaining essential features. The second layer expands the channel dimension to 128 and outputs $[128, 56, 56]$, capturing more complex spatial dependencies. The third layer further increases representational depth to $[256, 28, 28]$, focusing on semantically meaningful structures relevant to beam prediction. Finally, the fourth layer extracts high-level semantic features of size $[512, 14, 14]$, where spatial resolution is reduced while abstract information density increases. A final flattened representation of these feature maps is passed through a fully connected layer to yield a 512-dimensional visual embedding, which is used for subsequent multi-modal fusion.
\subsection{Transformer Feature Extraction}
As illustrated in Fig.~\ref{fig:2}, the Transformer module is designed to process structured sensor data and capture the intrinsic dependencies among multiple motion and position features. Although the structured input vector has a relatively small dimension, it contains heterogeneous information from GPS and IMU sensors, including UAV position, velocity, and attitude angles. Modeling the interactions among these heterogeneous attributes is crucial for accurate beam prediction in dynamic low-altitude environments.

The input structured feature $\hat{\mathbf{S}}[t] \in \mathbb{R}^{1 \times 8}$ is first projected into a latent space through a linear transformation to obtain $\mathbf{H}_{\text{struct}}^0 \in \mathbb{R}^{1 \times 512}$, followed by two Transformer encoder layers. Every encoder layer in this sequence employs a multi-head self-attention mechanism to characterize cross-dimensional dependencies and a feed-forward network to capture nonlinear interactions. Specifically, for each attention head in layer $l$ with input $\mathbf{H}_{\text{struct}}^l$, matrices for query, key, and value are calculated as:
\begin{equation}
    \mathbf{Q} = \mathbf{H}_{\text{struct}}^l\mathbf{W}_Q, \quad
    \mathbf{K} = \mathbf{H}_{\text{struct}}^l\mathbf{W}_K, \quad
    \mathbf{V} = \mathbf{H}_{\text{struct}}^l\mathbf{W}_V,
    \label{eq11}
\end{equation}
and the attention output for each head is given by:
\begin{equation}
    \text{head}_i = \text{Attention}(\mathbf{Q}, \mathbf{K}, \mathbf{V})
    = \text{softmax}\left(\frac{\mathbf{Q}\mathbf{K}^\top}{\sqrt{d_k}}\right)\mathbf{V},
    \label{eq12}
\end{equation}
where $d_k$ denotes the feature dimension of each head. Through multi-head attention, the model is able to jointly focus on information from various subspaces, and linearly projecting the concatenated outputs of all heads gives:
\begin{equation}
    \text{MultiHead}(\mathbf{H}_{\text{struct}}^l) = \text{Concat}(\text{head}_1, \dots, \text{head}_8)\mathbf{W}_O,
    \label{eq13}
\end{equation}
where $\mathbf{W}_O$ is the output projection matrix. Layer normalization and residual connections get applied to each sub-layer for stabilizing training and improving generalization.

Through this structure, the Transformer learns implicit relationships among position, velocity, and attitude dimensions, producing a refined structured embedding $\mathbf{F}_{\text{struct}} \in \mathbb{R}^{1 \times 512}$. This embedding is subsequently passed into the cross-attention fusion module, where it interacts with the visual features $\mathbf{F}_{\text{img}}$ to generate the fused multimodal representation $\mathbf{F}_{\text{fused}}$.
\subsection{Cross-Attention Fusion and Classification Module}
\begin{figure}[t]
    \centering
    \includegraphics[width=0.5\textwidth]{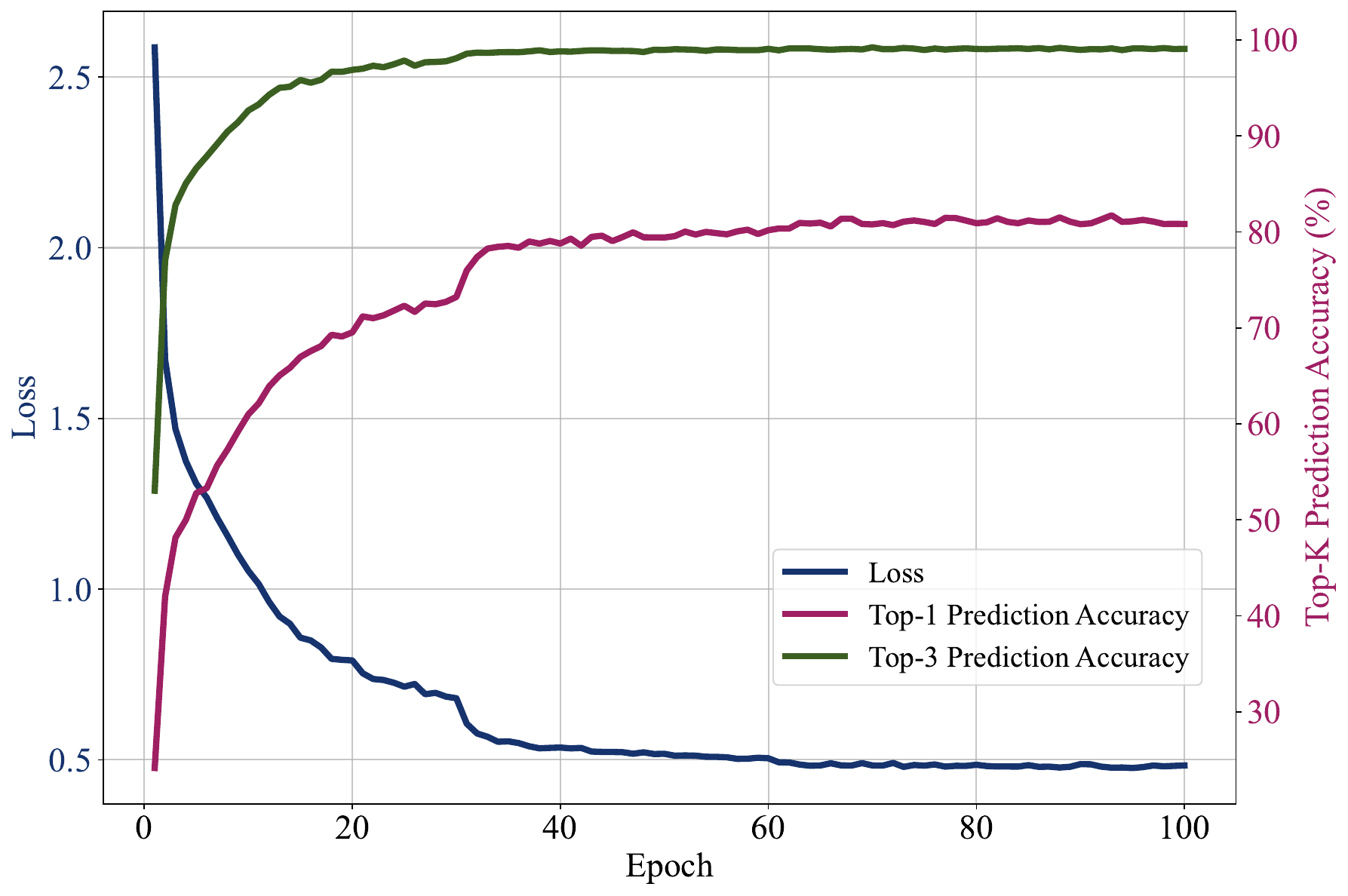}
    \caption{Prediction accuracy comparison under different learning rates. The learning rate of \(1\times10^{-4}\) reaches the greatest Top-1 and Top-3 accuracy, while higher (\(1\times10^{-3}\)) or lower (\(1\times10^{-5}\)) rates lead to slower or unstable convergence.}
    \label{fig:4}
\end{figure}
\subsubsection{Cross-Attention Fusion}
The cross-attention fusion module enables deep interaction between visual and structured modalities, surpassing simple concatenation. Visual data capture environmental context, while sensor data provide precise positional cues; these complementary modalities are adaptively integrated through cross-attention to learn interdependencies and emphasize relevant features. Given the visual embedding $\mathbf{F}_{\text{img}} \in \mathbb{R}^{1 \times 512}$ and structured embedding $\mathbf{F}_{\text{struct}} \in \mathbb{R}^{1 \times 512}$, we employ a multi-head cross-attention mechanism following the same formulation as in Eqs.~(\ref{eq11})--(\ref{eq13}). The key distinction lies in the input sources: visual features $\mathbf{F}_{\text{img}}$ serve as queries, while structured features $\mathbf{F}_{\text{struct}}$ provide keys and values. This design enables the network to refine visual representations based on sensor-derived spatial awareness, effectively aligning environmental context with UAV state information.

The cross-attention output $\mathbf{O}$ is combined with the original image features via residual connection and layer normalization:
\begin{equation}
\mathbf{F}'_{\text{img}} = \text{LayerNorm}(\mathbf{F}_{\text{img}} + \mathbf{O}),
\end{equation}
resulting in sensor-aware visual features that are both semantically enhanced and physically grounded. The enhanced visual features are then concatenated with the structured features along the feature dimension:
\begin{equation}
\mathbf{F}_{\text{fused}} = \text{Concat}(\mathbf{F}'_{\text{img}}, \mathbf{F}_{\text{struct}}),
\end{equation}
forming a fused multimodal representation $\mathbf{F}{\text{fused}} \in \mathbb{R}^{1 \times 1024}$ that encapsulates both environmental perception and motion context. A lightweight feed-forward network with residual connections is subsequently applied to further integrate these modalities and suppress redundant information.
\subsubsection{Classifier Design}
The fused representation $\mathbf{F}_{\text{fused}}$ is processed by a two-layer classifier that maps it to the $Q=64$ beam indices in codebook $\mathcal{F}$. The first layer projects the 1024-dimensional features into a latent space with ReLU and dropout, while the second outputs logits for beam prediction. This yields the predicted index, identifying the optimal beam while balancing expressiveness and efficiency for dynamic UAV environments.
\section{Experimental Results}
\begin{figure}[t]
    \centering
    \includegraphics[width=0.5\textwidth]{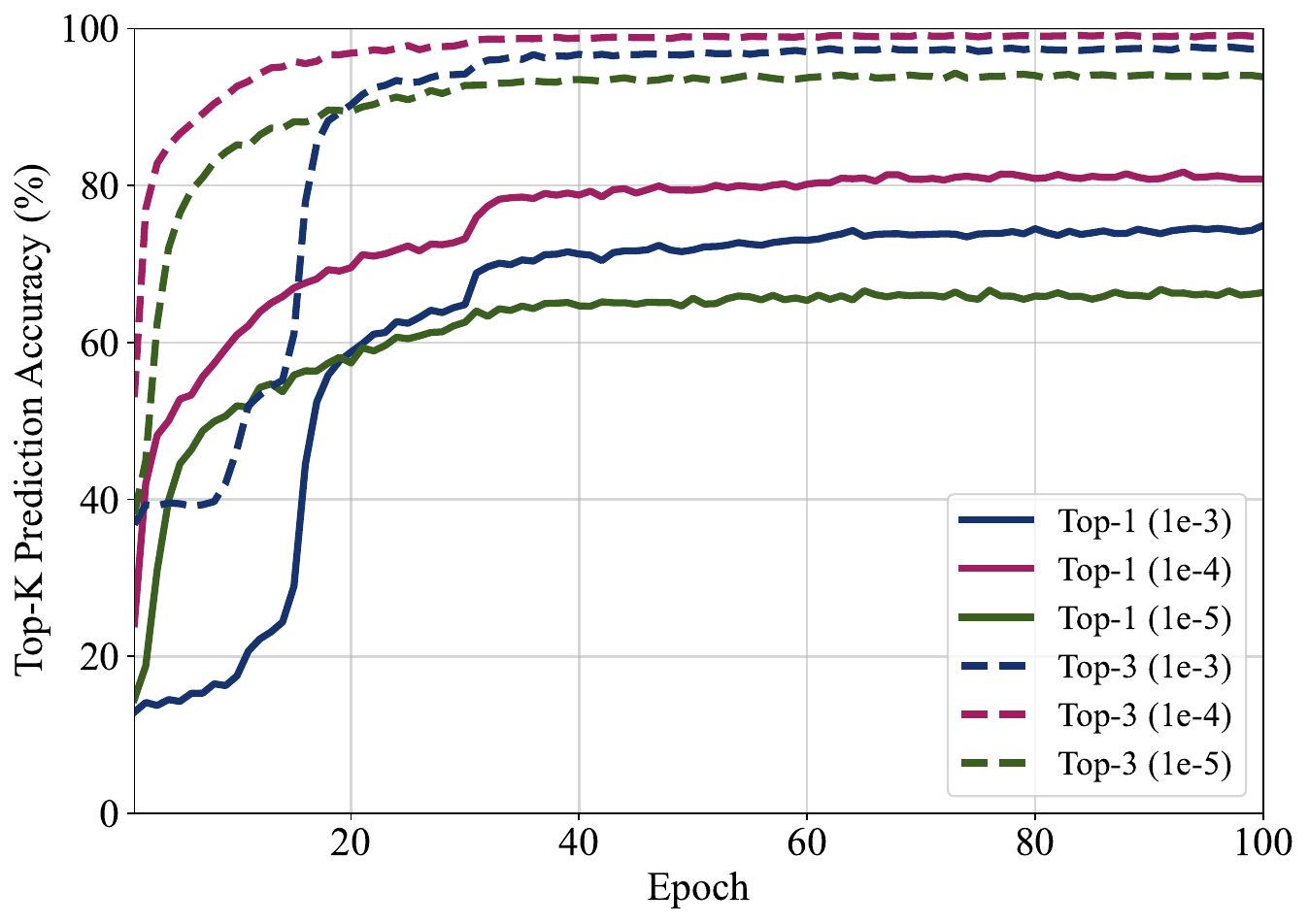}
    \caption{Prediction accuracy comparison under different learning rates.}
    \label{fig:5}
\end{figure}
\subsection{Dataset and Configuration}
For performance evaluation, we validate the proposed model on DeepSense 6G's Scenario 23 dataset \cite{deepsense6G} (Fig.~\ref{fig:3}). It has 11,387 samples: RGB camera visuals from a fixed BS, UAV multi-modal data (GPS, altitude, 3D velocity, attitude angles), and ground-truth optimal beam indices. Experiments use PyTorch on an NVIDIA RTX 4090 GPU. Training: 100 epochs, batch size 32, Adam (initial learning rate \(1\times10^{-4}\), step decay at 30/60/90 epochs). Dataset split: 80\% training, 20\% testing; CrossEntropyLoss as objective.

\subsection{Performance Analysis}
As shown in Fig.~\ref{fig:4}, during the initial 20 epochs, the training loss decreases rapidly while both Top-1 and Top-3 accuracies improve sharply, indicating that the model effectively captures underlying multimodal correlations. After around 40 epochs, the training process reaches a stable convergence stage, and accuracy continues to improve at a slower rate. By epoch 80, the model stabilizes with a Top-1 accuracy near 80\% and a Top-3 accuracy approaching 100\%, indicating consistent convergence and strong generalization.

This training behavior highlights the structural advantages of the proposed model. The CNN branch efficiently extracts hierarchical spatial features from visual data, the Transformer module captures interdependencies among structured sensor features, and the cross-attention fusion module enables deep integration between the two modalities. Together, these components allow synergistic exploitation of environmental and kinematic information, resulting in accurate beam prediction.

Fig.~\ref{fig:5} presents the performance comparison under different learning rates. The proposed model achieves the best performance when the learning rate is configured to \(1\times10^{-4}\), with a Top-1 accuracy exceeding 80\% and a Top-3 accuracy approaching 100\%. In contrast, higher (\(1\times10^{-3}\)) and lower (\(1\times10^{-5}\)) learning rates lead to suboptimal convergence, either due to instability or insufficient parameter updates. These results confirm that a \(1\times10^{-4}\) learning rate provides the greatest balance between convergence speed and training stability, ensuring effective optimization of multimodal features for robust beam prediction.

\begin{figure}[t]
    \centering
    \includegraphics[width=0.5\textwidth]{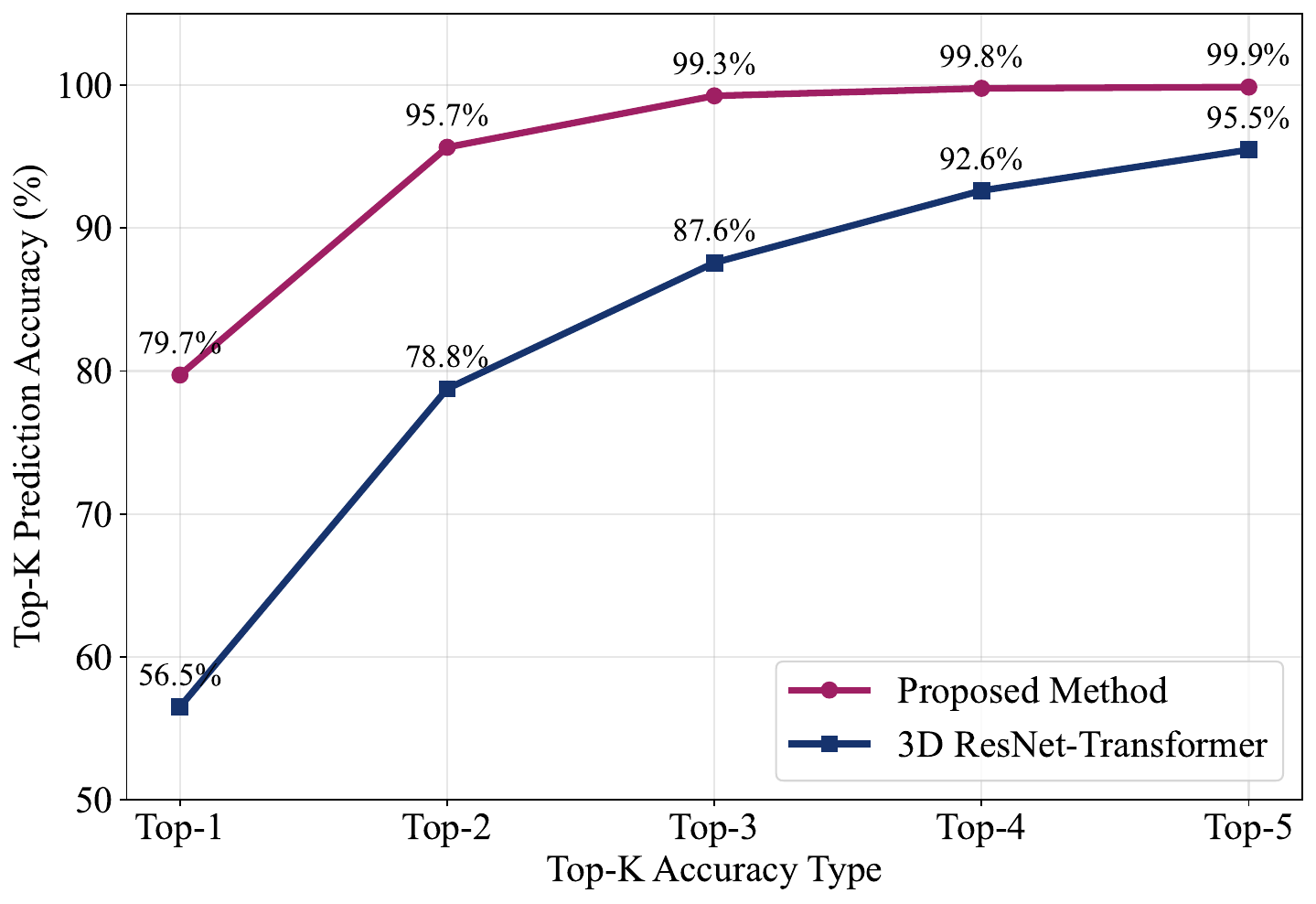}
    \caption{Comparison of performance between our proposed model and the 3D ResNet–Transformer baseline \cite{cui2024sensing}. The proposed cross-attention fusion method achieves significant improvements in Top-1, Top-3, and Top-5 accuracy, validating its superiority for low-altitude beam prediction.}
    \label{fig:6}
\end{figure}
A comprehensive contrast between the proposed model and the current 3D ResNet–Transformer fusion method \cite{cui2024sensing} is presented in Fig.~\ref{fig:6}. The proposed cross-attention-based fusion framework demonstrates substantial improvements across all Top-K accuracy metrics. Specifically, the Top-1 accuracy reaches 79.7\%, representing a 23.2\% improvement over the baseline. For Top-3 and Top-5 accuracies, the proposed model achieves 99.3\% and 99.9\%, surpassing the baseline by 11.7\% and 4.4\%, respectively.

These gains mainly result from the cross-attention fusion mechanism, which facilitates bidirectional feature interaction between image and sensor modalities rather than simple concatenation. By effectively modeling the underlying correlations between environmental perception and UAV kinematics, the proposed network achieves superior feature representation capability and prediction accuracy in dynamic low-altitude communication scenarios.

\subsection{Confusion Matrix Analysis}
\begin{figure}[t]
    \centering
    \includegraphics[width=0.44\textwidth]{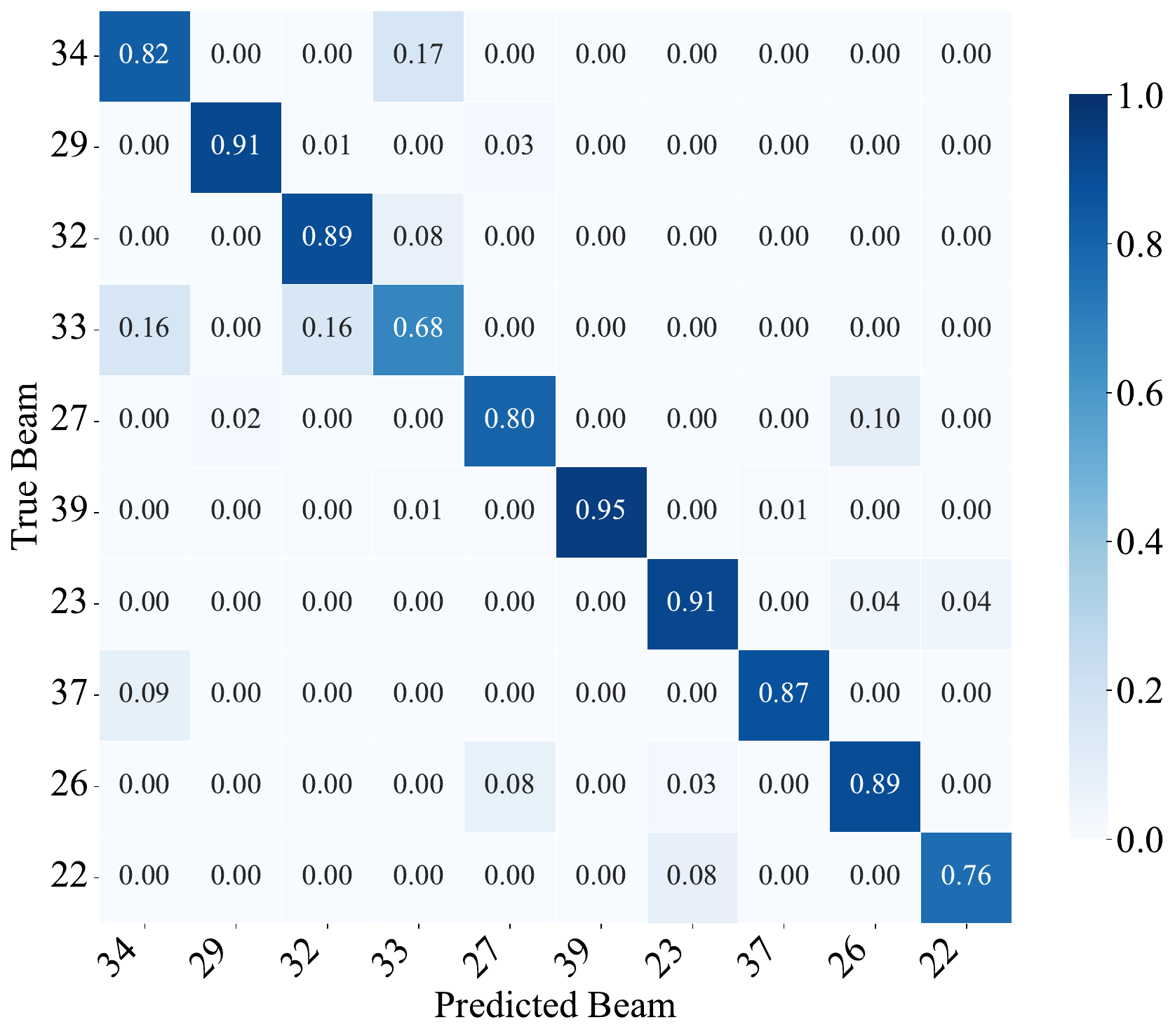}
    \caption{Confusion matrix of the Top 10 beam categories by sample count.}
    \label{fig:7}
\end{figure}

The classification performance across major beam categories is further illustrated in Fig.~\ref{fig:7}. Among the ten categories with the largest sample counts, the model achieves diagonal accuracies exceeding 80\% in most cases, with Beams 39, 29, and 23 reaching over 90\%. This indicates a strong ability to identify dominant beam patterns. Notably, the model maintains high discriminability even for physically similar beam pairs, such as Beam 34–Beam 33 (16.22\% confusion, 82.28\% accuracy) and Beam 27–Beam 26 (10.31\% confusion, 80.41\% accuracy). These results further confirm that the cross-attention fusion effectively integrates complementary visual and sensor cues, enabling robust distinction among highly correlated beam directions.
\section{Conclusion}
This paper presents a cross-attention based CNN-Transformer network for multi-modal beam prediction in low-altitude UAV networks. The proposed method achieves 79.7\% Top-1 and 99.3\% Top-3 accuracy, significantly outperforming existing approaches. By effectively fusing visual and sensor data through cross-attention, our solution provides reliable beam management for dynamic UAV communications and lays the groundwork for future research.
\vspace{12pt}
\bibliographystyle{IEEEtran} 
\bibliography{IEEEabrv, ref} 
\end{document}